\documentclass[10pt,preprint]{aastex}
\newcommand{\bm}[1]{{\mbox{\boldmath$#1$}}}
\shorttitle{Cosmological perturbations}
\shortauthors{Kopeikin Sergei et al.}
\begin{document}
\title{Cosmological perturbations: a new gauge-invariant approach}
\author{Sergei M. Kopeikin\altaffilmark{1}, Juan Ramirez\altaffilmark{2},
Bahram Mashhoon\altaffilmark{1} and Mikhail V. Sazhin\altaffilmark{3}}
\altaffiltext{1}{Department of Physics and Astronomy, University of
Missouri-Columbia, Columbia, MO 65211, USA}
\altaffiltext{2}{Departamento de Fisica Teorica I, Universidad Complutense de Madrid, 28040 Madrid, Spain}
\altaffiltext{3}{Sternberg State Astronomical Institute, Moscow State University, V-234, 119899 Moscow, Russia}
\begin{abstract}\noindent
A new gauge-invariant approach for describing cosmological
perturbations is developed. It is based on a physically motivated splitting
of the stress-energy tensor of the perturbation into two parts --- the {\it bare}
perturbation and
the {\it complementary} perturbation 
associated with stresses in the background gravitational field
induced by the introduction of the {\it bare} perturbation. The {\it complementary}
perturbation of the stress-energy tensor is explicitly singled
out and taken to the left side of the perturbed Einstein equations
so that the {\it bare} stress-energy tensor is the sole
source for the perturbation of the metric tensor and both sides of these equations are gauge invariant with respect to infinitesimal coordinate
transformations. For simplicity
we analyze the perturbations of the spatially-flat Friedmann-Lema\^{\i}tre-Robertson-Walker (FLRW) dust model.
A cosmological gauge can be chosen such that the equations for the
perturbations of the metric tensor are completely decoupled for the
$h_{00}, h_{0i}$, and $h_{ij}$ metric components and explicitly
solvable in terms of retarded integrals.
\end{abstract}
\newpage
\section{Introduction}\noindent
The relativistic theory of perturbations in spatially homogeneous
and isotropic cosmological models was first developed by E.M.
Lifshitz \citep{1,lkh}. At present, two approaches are commonly
used for the analysis of cosmological perturbations: the first
approach is based on globally defined coordinates in the perturbed
universe \citep{1,lkh,sw, silk,fut,5} and the second one
\citep{2,kodsas,3,4,dur1,dur2,5}, generally known as the covariant
approach, does not invoke any global coordinates but is based on a
slicing and threading of the space-time with a specific choice of
an orthonormal frame at each point. The main difficulty
associated with the two approaches is that $\delta T_{\alpha\beta}$ 
has always been used as the physical source for the primordial
perturbation of the metric tensor $h_{\alpha\beta}$ via the linearized
Einstein equations. Such a direct interpretation of $\delta T_{\alpha\beta}$
does not completely elucidate the physics underlying the perturbation theory. In general $\delta T_{\alpha\beta}$ must depend not only on the perturbation of matter but on the metric tensor perturbation $h_{\alpha\beta}$  
as well. Nevertheless, the a priori dependence of $\delta T_{\alpha\beta}$ on $h_{\alpha\beta}$ has remained implicit in previous approaches.
It is the purpose of this paper to make this dependence explicit by developing a new approach to cosmological perturbation theory. 

The theory developed here is based on a physically motivated splitting
of the perturbed stress-energy tensor into two parts --- the {\it bare}
perturbation and
the {\it complementary} perturbation  
associated with stresses in the background gravitational field
induced by the introduction of the {\it bare} perturbation. The {\it complementary}
perturbation of the stress-energy tensor is explicitly singled
out and taken to the left side of the perturbed Einstein equations
so that the {\it bare} stress-energy tensor turns out to be the sole
source for the perturbation of the metric tensor. We also require that both sides of 
the linearized Einstein equations be independently invariant under gauge (Lie) transformations of the metric tensor perturbations. This approach allows us to find a new cosmological gauge such that the perturbation equations decouple and their solutions can be found explicitly in terms of retarded integrals.  

The development of the new approach was independently started in
papers by \cite{juan} and \cite{sergei}, where a new cosmological gauge in the
theory of cosmological perturbations was explored. Our aim here is to generalize the treatment started in these
papers for the case of the dust-dominated spatially-flat
FLRW cosmological model, which
is adopted here for the sake of simplicity. Possible extension to other models with different equations of state of the background matter can be pursued in a similar way.

We describe the background universe in Section 2. Then,
in Section 3 the gauge invariance and
integrability of the Einstein equations are used for a unique
scalar-tensor decomposition of the perturbed background
stress-energy tensor in such a way that the 
Einstein equations for the metric perturbations are explicitly
obtained with the {\it bare} stress-energy tensor of matter acting as the source of the perturbation of the metric tensor. A
new cosmological gauge is discussed in Section 4 for reducing the
gauge freedom of the linearized Einstein equations and bringing
them to the form of the wave-type equations which are explicitly
solved in Section 5 in the same way as 
d'Alembert's equation is solved in flat space-time.

In what follows we use the `geometrized' units in which $G=c=1$
and the conventions adopted in the textbook of \cite{6}.
Greek indices in the full Einstein equations are raised and lowered
with the help of the complete metric $g_{\alpha\beta}$ and those in
the linearized Einstein equations are raised
and lowered with the help of the background metric
$\overline{g}_{\alpha\beta}$. 
\section{The background universe}\noindent
Let us choose the background cosmological model to be the
spatially-flat FLRW space-time given by the
metric
\begin{equation}
\label{1}
\overline{g}_{\alpha\beta}=a^2(\eta)f_{\alpha\beta}\;,\qquad\qquad
\overline{g}^{\alpha\beta}={1\over a^{2}(\eta)}f^{\alpha\beta}\;,
\end{equation}
where hereafter the overbar refers to the background,
$f_{\alpha\beta}={\rm diag}(-1,+1,+1,+1)$ is the Minkowski
metric, $\eta$ is a dimensionless temporal coordinate related to the cosmic time
$t$ by the first-order ordinary differential equation
$dt=a(\eta)d\eta$, and $a(\eta)$ is a scale factor with the dimension of length. The cosmic time coincides with the proper time of static observers in the background space-time. The background
Einstein equations read
\begin{equation}
\label{2}
\overline{G}_{\alpha\beta}\equiv\overline{R}_{\alpha\beta}-{1\over
2}\overline{g}_{\alpha\beta}\overline{R}=8\pi\overline{T}_{\alpha\beta}\;,
\end{equation}
where $\overline{R}_{\alpha\beta}$ is the Ricci tensor,
$\overline{R}=\overline{R}^{\alpha}_{\;\alpha}$ is the scalar
curvature, 
$\overline{T}_{\alpha\beta}=\overline{\rho}\;\overline{u}_\alpha\overline{u}_\beta$
is the stress-energy tensor of the background matter, and
$\overline{\rho}(\eta)$  and $\overline{u}_\alpha(\eta)$ are the
density and four-velocity of the matter, respectively. The
background universe is spatially homogeneous and isotropic, hence,
$\overline{u}_\alpha(\eta)=a^{-1}(\eta)\overline{g}_{0\alpha}=-a(\eta)\delta^0_\alpha$.
Let us introduce the Hubble `parameter'
${H}(\eta)=\dot{a}(\eta)/a^2(\eta)$, where the
overdot denotes the time derivative with respect to $\eta$, so
that
\begin{equation}
\label{usef} \overline{R}_{\alpha\beta}={\dot{H}\over
a}\left(\overline{g}_{\alpha\beta}-2\overline{u}_\alpha\overline{u}_\beta\right)+
3H^2\overline{g}_{\alpha\beta}\;,\qquad\quad\overline{R}=6\left({\dot{H}\over
a}+2H^2\right)\;.
\end{equation}
Equations (\ref{1})--(\ref{usef}) have a unique solution \citep{ll}:
\begin{equation}
\label{3}
a(\eta)={2\eta^2\over {H}_0}\;,\qquad {H}(\eta)={{
H}_0\over\eta^3}\;,\qquad\overline{\rho}(\eta)={3{H}_0^2\over 8\pi\eta^6
}\;,\qquad t={2\eta^3\over 3{H}_0}\;,
\end{equation}
where $\eta\equiv\eta_0=1$ at the present epoch, and ${H}_0$ is the
present value of the Hubble parameter ${H}(\eta_0)$. It
is worth noting for the following calculations that
$\dot{H}=-(3/2)H^2 a$ and
$\overline{u}_{\alpha|\beta}=H P_{\alpha\beta}$, where
$P_{\alpha\beta}=\overline{u}_{\alpha}\overline{u}_{\beta}+\overline{g}_{\alpha\beta}$
is the projection tensor on the hypersurface orthogonal to the
unperturbed four-velocity $\overline{u}_{\alpha}$.
\section{Basic assumptions}\noindent
Let us assume that the background space-time metric
$\overline{g}_{\alpha\beta}$ is weakly perturbed by the presence
of a disturbance with the stress-energy tensor
$T^{(m)}_{\alpha\beta}$ of arbitrary origin (e.g., a galaxy,
a background fluid density perturbation, a cosmic string, etc.). This {\it bare} perturbation,
in the absence of any interaction with the background matter,
would be expected to move in the background space-time in such a
way that
\begin{equation}
\label{4}
T^{(m)\beta}_{\;\alpha\quad\;|\beta}=0\;,
\end{equation}
where the vertical bar denotes a covariant derivative with respect
to the background metric $\overline{g}_{\alpha\beta}$. The tensor
$T^{(m)}_{\alpha\beta}$ must be understood as a weak perturbation
of the background space-time metric only; however, equation
(\ref{4}) does not prohibit the energy density contrast
$T^{(m)}_{00}/\overline{T}_{00}$ to be very large in some local
region or even be singular. On the other hand, if one wants
to consider the evolution of primordial cosmological perturbations
that are not so well localized such as, e.g. a black hole, it must be
assumed that the averaged energy density contrast
$\langle T^{(m)}_{00}\rangle/\langle\overline{T}_{00}\rangle \ll 1$ in most regions of the
space-time manifold.

The presence of the {\it bare} perturbation implies that the full
space-time metric $g_{\alpha\beta}$ can be written as a linear sum
of the background metric $\overline{g}_{\alpha\beta}$ and the
space-time perturbation $h_{\alpha\beta}$
\begin{equation}
\label{5}
g_{\alpha\beta}(\eta, {\bm{\it x}})=\overline{g}_{\alpha\beta}+h_{\alpha\beta}(\eta, {\bm{\it x}})\;,
\end{equation}
where we have introduced a four-dimensional coordinate chart
$x^\alpha=(\eta, {\bm{\it x}})$ on the background space-time
manifold. In what follows it is convenient to introduce a new
variable defined as
\begin{equation}
\label{5a}
\psi_{\alpha\beta}=h_{\alpha\beta}-{1\over 2}\;\overline{g}_{\alpha\beta}\;h\;,
\end{equation}
where $h\equiv h^\alpha_{\;\alpha}=
\overline{g}^{\alpha\beta}h_{\alpha\beta}=-\psi=-\overline{g}^{\alpha\beta}\psi_{\alpha\beta}$. The Einstein equations for the full metric read
\begin{equation}
\label{6}
G_{\alpha\beta}\equiv R_{\alpha\beta}-{1\over 2}g_{\alpha\beta}R=8\pi T_{\alpha\beta}\;.
\end{equation}
These equations can be linearized after making use of equations
(\ref{2}) and (\ref{5}). The result is
\begin{equation}
\label{6a}
G_{\alpha\beta}=\overline{G}_{\alpha\beta}+\delta G_{\alpha\beta}\;,\qquad\quad
T_{\alpha\beta}=\overline{T}_{\alpha\beta}+\delta T_{\alpha\beta}\;.
\end{equation}
Here $\delta
T_{\alpha\beta}$ is a perturbation of the background stress-energy
tensor and $\delta G_{\alpha\beta}$ is a perturbation of the Einstein
tensor that in the case of homogeneous and
isotropic background space-time can be written as
\begin{equation}
\label{6b} \delta G_{\alpha\beta}=-{1\over
2}\left({\psi_{\alpha\beta}}^{|\nu}_{\;\;\;|\nu}+
\overline{g}_{\alpha\beta}B^\nu_{\;\;|\nu}-B_{\alpha|\beta}-B_{\beta|\alpha}\right)+
2\overline{R}^{\;\nu}_{\;(\alpha}\psi^{}_{\beta)\nu}-\frac{2}{3}\overline{R}\psi_{\alpha\beta}
-{1\over
2}\left(\overline{R}_{\alpha\beta}-\frac{1}{3}\overline{g}_{\alpha\beta}\overline{R}\right)\psi\;,
\end{equation}
where $B_\alpha\equiv {\psi_{\alpha}^{\;\;\nu}}_{|\nu}$. 
Assuming that the background Einstein equations (\ref{2})
are valid, the linearized Einstein equations read
\begin{equation}
\label{7} \delta G_{\alpha\beta}=8\pi\;\delta T_{\alpha\beta}\;.
\end{equation}

It is worth noting that the {\it bare} perturbation $T^{(m)}_{\alpha\beta}$
interacts gravitationally with the background space-time causing perturbations both of the background geometry and the background stress-energy tensor. As a result of this
gravitational interaction a complementary stress-energy tensor
$T^{(c)}_{\alpha\beta}$ is induced which explicitly depends on the metric perturbation $h_{\alpha\beta}$. For this reason in a linear
approximation, the {\it dressed} perturbation of the background
stress-energy tensor $\delta T_{\alpha\beta}$ can be written as a
linear sum
\begin{equation}\label{8}
\delta T_{\alpha\beta}=T^{(m)}_{\alpha\beta}+T^{(c)}_{\alpha\beta}\;.
\end{equation}
Because the {\it complementary} perturbation depends on $h_{\alpha\beta}$ we can take $T^{(c)}_{\alpha\beta}$ to the left side of the linearized Einstein equations (\ref{7}).
The {\it bare} perturbation remains in the right side of equation (\ref{7}) and has now a clear physical meaning: its
introduction into the background manifold originates the deviation
from the background metric $\overline{g}_{\alpha\beta}$ and leads
to the perturbed space-time manifold. The {\it complementary}
perturbation of the background stress-energy tensor is also
generated by the {\it bare} perturbation and, hence, can be represented as
\begin{equation}\label{9}
T^{(c)}_{\alpha\beta}=
\overline{u}_\alpha\overline{u}_\beta(\delta\rho+\delta p)+
\overline{\rho}\;
\overline{u}_\alpha\delta{u}_\beta+\overline{\rho}\;\overline{u}_\beta\delta{u}_\alpha+\overline{g}_{\alpha\beta}\delta p\;,
\end{equation}
where $\delta\rho$, $\delta p$, and $\delta u_\alpha$ are as yet
undetermined perturbations of the background energy density,
pressure, and the
four-velocity respectively. It is worth noting that though the background pressure in the dust-dominated cosmological model is identically equal to zero, its
perturbation $\delta p$ would be associated with the stresses of the background gravitational field and could be different from zero; that is, the {\it bare} perturbation disturbs the Hubble flow thereby inducing a purely gravitational effective pressure $\delta p$.

The structure of these perturbations
and, as a consequence, that of the complementary tensor
$T^{(c)}_{\alpha\beta}$ is completely determined by two main
constraints. The first constraint comes from the integrability condition for
equation (\ref{7}) that is a direct consequence of the linearized
Bianchi identity for the full Einstein equations (\ref{6}). Accounting for equation (\ref{4}) in linear approximation the Bianchi identity reads
\begin{equation}
\label{10}
\hat{T}_{\alpha}^{(c)\nu}{}_{|\nu}+\overline{T}^{\beta}_{\;\alpha}\delta{\Gamma}^{\nu}_{\beta\nu}
-\overline{T}^{\beta}_{\;\nu}\delta{\Gamma}^{\nu}_{\alpha\beta}=0\;,
\end{equation}
where $\hat{T}_{\alpha}^{(c)\nu}\equiv\overline{g}^{\nu\beta}T^{(c)}_{\alpha\beta}-h^{\nu\beta}\overline{T}_{\alpha\beta}$ and
$\delta{\Gamma}^{\nu}_{\alpha\beta}$ are perturbations of the Christoffel symbols
\begin{equation}
\label{11}
\delta{\Gamma}^{\nu}_{\alpha\beta}={1\over 2}\left(h^\nu_{\;\alpha|\beta}+h^\nu_{\;\beta|\alpha}-h_{\alpha\beta}^{\quad|\nu}\right)\;.
\end{equation}
The second constraint on the structure of $T^{(c)}_{\alpha\beta}$
comes from the gauge invariance of the Einstein equations and the {\it bare} perturbation. That is, if one chooses a slightly different coordinate chart
\begin{equation}
\label{ch}
x'^\alpha=x^\alpha-\xi^\alpha(\eta,{\bm{\it x}})\;,
\end{equation}
the gauge (Lie) transformations 
\begin{equation}
\label{lie}
\delta G'_{\alpha\beta}(\eta,{\bm{\it x}})=\delta G_{\alpha\beta}(\eta,{\bm{\it x}})+
{\cal L}_{\bm{\it\xi}}\overline{G}_{\alpha\beta}(\eta)\;,\qquad\quad
\delta T'_{\alpha\beta}(\eta,{\bm{\it x}})=\delta T_{\alpha\beta}(\eta,{\bm{\it x}})+
{\cal L}_{\bm{\it\xi}}\overline{T}_{\alpha\beta}(\eta)\;,
\end{equation}
where ${\cal L}_{\bm{\it\xi}}$ denotes a Lie derivative along the vector field ${\bm{\it\xi}}$, must preserve the form of the linearized Einstein equations (\ref{7})
\begin{equation}
\label{pp}
\delta G'_{\alpha\beta}(\eta,{\bm{\it x}})=8\pi\;\delta T'_{\alpha\beta}(\eta,{\bm{\it x}})\;.
\end{equation}
Here $\delta G'_{\alpha\beta}$ and $\delta T'_{\alpha\beta}$ are respectively defined by exactly the same relations as $\delta G_{\alpha\beta}$ and $\delta T_{\alpha\beta}$ after making the replacement $h_{\alpha\beta}\rightarrow h'_{\alpha\beta}=h_{\alpha\beta}+\xi_{\alpha|\beta}+\xi_{\beta|\alpha}$.
We emphasize that in the linear approximation the {\it bare} stress-energy tensor remains invariant under the gauge transformations (\ref{ch}), i.e. $T'^{(m)}_{\alpha\beta}(\eta,{\bm{\it x}})=T^{(m)}_{\alpha\beta}(\eta,{\bm{\it x}})$, since in our approach this tensor is the source for the perturbations of both the background geometry and the background stress-energy tensor. Hence, we conclude that in the new gauge $T'^{(c)}_{\alpha\beta}(\eta,{\bm{\it x}})=T^{(c)}_{\alpha\beta}(\eta,{\bm{\it x}})+
{\cal L}_{\bm{\it\xi}}\overline{T}_{\alpha\beta}(\eta)$.
We can now use the results of the present section to determine the structure of the {\it complementary} stress-energy tensor. 
\section{Gauge-invariant structure of the {\bm{\it complementary}} stress-energy tensor}\noindent
The second constraint on the structure of the $T^{(c)}_{\alpha\beta}$, discussed in Section 3, requires the introduction of supplementary fields $\Phi(\eta,{\bm{\it x}})$ and $Z_\alpha(\eta,{\bm{\it x}})$. We assume in analogy with equation (\ref{5}) for the metric perturbations that these scalar and vector fields can be expanded around their background values 
\begin{equation}
\label{rr}
\Phi=\overline{\phi}(\eta)+\phi(\eta,{\bm{\it x}})\;,\qquad\quad Z_\alpha=\overline{\zeta}_\alpha(\eta)+\zeta_\alpha(\eta,{\bm{\it x}})\;.
\end{equation}
Generalizing ideas developed in \cite{1} and \citep{2,kodsas,3,4}, we assume that the background matter perturbations can be linearly expressed in terms of the tensor field $\psi_{\alpha\beta}(\eta,{\bm{\it x}})$, the vector field $\zeta_\alpha(\eta,{\bm{\it x}})$, and the scalar field $\phi(\eta,{\bm{\it x}})$ with coefficients that are proportional to the background quantities depending only on time $\eta$. We demand that under the Lie transformations \citep{wei}
\begin{eqnarray}
\label{12}
\psi'_{\alpha\beta}(\eta,{\bm{\it x}})&=&\psi_{\alpha\beta}(\eta,{\bm{\it x}})+\xi_{\alpha|\beta}+\xi_{\beta|\alpha}-\overline{g}_{\alpha\beta}\xi^{\mu}_{\;\;|\mu}\;,\\\label{12a}
\zeta'_\alpha(\eta,{\bm{\it x}})&=&\zeta_\alpha(\eta,{\bm{\it x}})+\overline{\zeta}_{\alpha|\mu}\xi^{\mu}+\overline{\zeta}_\mu\xi^{\mu}_{\;\;|\alpha}\;,\\\label{12b}
\phi'(\eta,{\bm{\it x}})&=&\phi(\eta,{\bm{\it x}})+\overline{\phi}_{|\mu}\xi^\mu\;,
\end{eqnarray}
our linear expressions for the matter perturbations in terms of $\psi_{\alpha\beta}$, $\zeta_\alpha$, and $\phi$ must change in accordance with the general gauge transformations  
\begin{eqnarray}
\label{12c}
\delta\rho'(\eta,{\bm{\it x}})&=&\delta\rho(\eta,{\bm{\it x}})+\overline{\rho}_{|\mu}\xi^\mu\;,\\\label{12d}
\delta p'(\eta,{\bm{\it x}})&=&\delta p(\eta,{\bm{\it x}})\;,\\\label{12e}
\delta u'_\alpha(\eta,{\bm{\it x}})&=&\delta u_\alpha(\eta,{\bm{\it x}})+\overline{u}_{\alpha|\mu}\xi^{\mu}+\overline{u}_\mu\xi^{\mu}_{\;|\alpha}\;,
\end{eqnarray}
for our dust model.

Straightforward calculations based on these consistency requirements reveal that one can
choose the gauge vector field $Z_\alpha\equiv 0$ and $\overline{\phi}\equiv -3\ln a(\eta)$, so that $\overline{\phi}_{|\alpha}(\eta)=3{H}\overline{u}_\alpha$. Furthermore, 
the matter
perturbations can then be represented in the following scalar-tensor
form
\begin{eqnarray}
\label{19}
\delta\rho&=&\;\;\;{1\over 2}\left(\overline{T}^{\mu\nu}\psi_{\mu\nu}-{1\over 2}\;\overline{T}\psi\right)-{1\over 2}\overline{T}\phi-{H\over 8\pi}\overline{u}^\mu\;\phi_{|\mu}\;,\\\nonumber\\\label{20}
\delta p&=&\;\;\;{1\over 2}\left(\overline{T}^{\mu\nu}\psi_{\mu\nu}-{1\over 2}\;\overline{T}\psi\right)+{1\over 2}\overline{T}\phi-{H\over 8\pi}\overline{u}^\mu\;\phi_{|\mu}\;,\\\nonumber\\\label{21}
\overline{\rho}\;\delta u_\alpha&=&-{1\over 2}\left(\overline{T}^{\mu\nu}\psi_{\mu\nu}-{1\over 2}\;\overline{T}\psi\right)\overline{u}_\alpha+{H\over 8\pi}\left(\phi_{|\alpha}+\overline{u}_\alpha\overline{u}^\mu\;\phi_{|\mu}\right)\;.
\end{eqnarray}
The {\it complementary} stress-energy tensor $T^{(c)}_{\alpha\beta}$ is now determined by equations (\ref{9}) and (\ref{19})--(\ref{21}), and is given  as a sum of two pieces depending separately on the metric tensor perturbation $\psi_{\alpha\beta}$ and the scalar perturbation $\phi$,
\begin{eqnarray}
\label{15}
T^{(c)}_{\alpha\beta}&=&T^{(\psi)}_{\alpha\beta}+T^{(\phi)}_{\alpha\beta}\;,\\
\nonumber\\\label{16}
T^{(\psi)}_{\alpha\beta}&=&\frac{1}{2}\;\overline{g}_{\alpha\beta}
\left(\overline{T}^{\mu\nu}\psi_{\mu\nu}-\frac{1}{2}\overline{T}\;\psi\right)\;,\\
\nonumber\\\label{17}
T^{(\phi)}_{\alpha\beta}&=&\frac{1}{2}\;\overline{g}_{\alpha\beta}\;\overline{T}\;\phi+
{{ H}\over 8\pi}\left(\overline{u}_\alpha\;\phi_{\mid\beta}+\overline{u}_\beta\;\phi_{\mid\alpha}-\overline{g}_{\alpha\beta}\;\overline{u}^{\mu}\phi_{\mid\mu}\right)\;.
\end{eqnarray}
The gauge transformation of the complementary tensor is given by
\begin{equation}
\label{13}
T'^{(c)}_{\alpha\beta}(\eta,{\bm{\it x}})=T^{(c)}_{\alpha\beta}(\eta,{\bm{\it x}})+{\cal L}_{\bm\xi}\overline{T}_{\alpha\beta}(\eta)\;,
\end{equation}
where
\begin{equation}
\label{14}
{\cal L}_{\bm{\xi}}\overline{T}_{\alpha\beta}(\eta)=\overline{T}_{\alpha\beta|\nu}\;\xi^\nu+\overline{T}_{\alpha\nu}\;\xi^\nu_{\;\;|\beta}+\overline{T}_{\beta\nu}\;\xi^\nu_{\;\;|\alpha}\;
\end{equation}
is a Lie derivative of the background stress-energy tensor along
the vector field $\xi^\alpha$ that is generated by the coordinate
transformation (\ref{ch}). Taking into account the gauge
transformation equations (\ref{lie}) and (\ref{pp}), and the fact that
in the linear theory the {\it bare} stress-energy tensor
$T^{(m)}_{\alpha\beta}$ does not change under the gauge
transformations (\ref{12})--(\ref{12b}), we conclude that the
linearized Einstein equations can be written in an arbitrary
coordinate system as 
\begin{equation}
\label{eeq}
\delta G_{\alpha\beta}-8\pi T^{(c)}_{\alpha\beta}=8\pi T^{(m)}_{\alpha\beta}\;,
\end{equation}
where $\delta G_{\alpha\beta}$ is defined by equation (\ref{6b}).

The supplementary scalar field $\phi$ is determined by making use of the first constraint on $T^{(c)}_{\alpha\beta}$, i.e.
the integrability condition (\ref{10}), which gives
a specific equation relating the vector functions $B_\alpha$ and
the scalar field $\phi$; this equation is invariant with
respect to the gauge transformations (\ref{12})--(\ref{12b}) and
has the following form
\begin{equation}
\label{18a}
\overline{u}^\alpha B_\alpha=-{5\over 2}{H}\;\overline{u}^\alpha\overline{u}^\beta\psi_{\alpha\beta}-{1\over 4}{H}\psi-{3\over 2}{H}\phi+{1\over 3{H}}\phi^{|\alpha}_{\;\;\;|\alpha}\;.
\end{equation}
We interpret this relation as the second-order differential equation for the scalar field $\phi$ whose solutions depend on our choice of the (gauge) vector functions $B_\alpha$.

The difference between our treatment of the perturbation of the stress-energy tensor $\delta T_{\alpha\beta}$ and the gauge-invariant approach adopted by previous authors \citep{2,kodsas,3,4} is that we clearly distinguish the a priori {\it bare} perturbation from the {\it complementary} perturbation induced by the introduction of the {\it bare} perturbation into the background space-time. In addition, we are able to represent the {\it complementary} perturbation $T^{(c)}_{\alpha\beta}$ of the background stress-energy tensor as an explicit function of the tensor and scalar perturbations that obey well-defined hyperbolic differential equations. 

Our approach to cosmological perturbations appears to be rather suitable for situations where the bare perturbation can be thought of as an isolated physical system, since for such systems $T^{(m)}_{\alpha\beta}$ can be immediately determined explicitly as, e.g., in the case of cosmic strings or domain walls \citep{kibble, string}. It would be interesting to explore in detail the relationship between our treatment and previous approaches. This is beyond the scope of the present endeavor, however, and will be dealt with in a future study. Here we only delineate some of the novel features of our formalism. 

\section{The gauge-invariant linearized Einstein equations for gravitational perturbations}\noindent
Making use of formulas (\ref{15})--(\ref{17}) for the {\it complementary} tensor $T^{(c)}_{\alpha\beta}$, one can
transform the linearized Einstein equations (\ref{eeq}) to the form
\begin{eqnarray}
\nonumber
{\psi_{\alpha\beta}}^{|\nu}_{\;\;\;|\nu}&+&\overline{g}_{\alpha\beta}B^\nu_{\;\;|\nu}-B_{\alpha|\beta}-B_{\beta|\alpha}-4\overline{R}^{\;\nu}_{\;(\alpha}\psi^{}_{\beta)\nu}+\frac{4}{3}\;\overline{R}\;\psi_{\alpha\beta}
+\left(\overline{R}_{\alpha\beta}-\frac{1}{3}\overline{g}_{\alpha\beta}\overline{R}\right)\psi\\\nonumber\\&+&\overline{g}_{\alpha\beta}\left(\overline{R}^{\mu\nu}\psi_{\mu\nu}-\overline{R}\;\phi\right)+2{H}\left(\overline{u}_\alpha\;\phi_{\mid\beta}+\overline{u}_\beta\;\phi_{\mid\alpha}-\overline{g}_{\alpha\beta}\;\overline{u}^{\mu}\phi_{\mid\mu}\right)
=-16\pi\; T^{(m)}_{\alpha\beta}\;,\label{22}
\end{eqnarray}
such that the left side is gauge invariant. This can be proved by
applying the gauge transformations (\ref{12})--(\ref{12b}) to the left side of Einstein's equations (\ref{22}) and accounting for the fact that the {\it bare} stress-energy tensor is gauge invariant by definition. Equation (\ref{18a}) for the scalar field $\phi$ can be re-written in the source-free form
\begin{equation}
\label{phi}
\phi^{|\alpha}_{\;\;\;|\alpha}-{3\over 2}{H}^2\left(3\phi+{1\over 2}\psi+5\overline{u}^\alpha\overline{u}^\beta\psi_{\alpha\beta}\right)-3{H}\overline{u}^\alpha B_\alpha=0\;.
\end{equation}
This equation is also invariant under the gauge transformations (\ref{12})--(\ref{12b}).

The set of equations (\ref{22})--(\ref{phi}) constitutes the basis for the new gauge-invariant approach to cosmological perturbations of the dust-dominated FLRW model. We have weighty arguments that an analogous approach can be worked out for the more general `canonical' equation of state of the background matter $\overline{p}=\alpha\overline{\rho}$ with arbitrary numerical value of $\alpha$ as well as for the case of a universe with $\Lambda$-term and a background fluid with components that have different equations of state. 

The vector field $B_\alpha$ and the scalar field
$\phi$ are arbitrary gauge functions related through equation (\ref{phi}). A specific choice of these functions restricts the gauge freedom and can simplify the linearized Einstein equations. After choosing a specific gauge, the residual coordinate freedom is defined by the functions $\xi^\alpha$ describing the gauge transformations (\ref{ch}). These functions obey the following inhomogeneous equation
\begin{equation}
\label{gg}
\xi^{\alpha|\beta}_{\quad|\beta}+\overline{R}^\alpha_{\;\beta}\;\xi^\beta=B'^\alpha(\eta,{\bm{\it x}})-B^\alpha(\eta,{\bm{\it x}})\;,
\end{equation}
that is obtained by differentiation of equation (\ref{12}) and making use of the definitions $B'_\alpha\equiv\psi'^\beta_{\;\;\alpha|\beta}$ and $B_\alpha\equiv\psi^\beta_{\;\alpha|\beta}$. Once $B_\alpha$ is specified, four restrictions are imposed on the choice of the coordinate system. Then, equations (\ref{gg}) describe four residual degrees of gauge freedom in this coordinate system. Thus, the total number of functional restrictions on the eleven degrees of freedom of the gravitational field $\psi_{\alpha\beta}$ and the scalar field $\phi$ is eight, which means that we have three independent variables in the case of a free gravitational field propagating on the curved background instead of two that would be expected in the case of asymptotically-flat space-time. Two of the three degrees of freedom describe $\otimes$ and $\oplus$ polarizations of gravitational waves and the third degree of freedom belongs to the scalar field $\phi$.
\section{New cosmological gauge}\noindent
Our new gauge-invariant approach to the theory of cosmological perturbations allows us to decouple the linearized Einstein equations. This simplifies the task of finding their solutions. Such a decoupling can be achieved in the framework of a new cosmological gauge that is defined by the condition
\begin{equation}
\label{gc}
\phi_{|\alpha}={3\over 2}{H}\left(\overline{u}_\alpha\phi+{1\over 2}\overline{u}^\beta\psi_{\alpha\beta}\right)\;,
\end{equation}
leading to the second-order hyperbolic equation for the scalar field $\phi$,
\begin{equation}
\label{eqphi}
\phi^{|\alpha}_{\;\;\;|\alpha}+{3\over 2}{H}^2\left(\phi-\overline{u}^\alpha\overline{u}^\beta\psi_{\alpha\beta}-{1\over 2}\psi\right)=0\;,
\end{equation}
and a particular solution of equation (\ref{phi}) for the vector field $B_\alpha$,
\begin{equation}
\label{bbz}
B_\alpha=2{H}\left(\overline{u}_\alpha\phi-\overline{u}^\beta\psi_{\alpha\beta}\right)\;.
\end{equation}
It is convenient to introduce the new metric variables $\varphi_{\alpha\beta}$: 
\begin{equation}
\label{24}
\psi_{\alpha\beta}(\eta,{\bm{\it x}})=a^2(\eta)\;\varphi_{\alpha\beta}(\eta,{\bm{\it x}})\quad\;,\quad \varphi(\eta,{\bm{\it x}})=f^{\alpha\beta}\varphi_{\alpha\beta}(\eta,{\bm{\it x}})\quad\;.
\end{equation}
Then, Einstein's equations (\ref{22}), in the gauge defined by equation (\ref{bbz}), can be reduced to the following form
\begin{equation}
\label{dec}
\Box\varphi_{\alpha\beta}-2{\cal H}\varphi_{\alpha\beta},_0+{\cal H}^2\left[\delta^0_\alpha\varphi_{\beta 0}+\delta^0_\beta\varphi_{\alpha 0}+\delta^0_\alpha\delta^0_\beta(\varphi-2\phi)\right]=-16\pi\; T^{(m)}_{\alpha\beta}\;,
\end{equation}
where $\Box\equiv f^{\alpha\beta}\partial_\alpha\partial_\beta$, and ${\cal H}=\dot{a}/a$.
Equations (\ref{eqphi}) and (\ref{dec}) are equivalent to the following  set of inhomogeneous hyperbolic equations
\begin{eqnarray}
\label{25q}
\Box \chi-\frac{4}{\eta}{\partial \chi\over\partial\eta}+\;{10\chi\over\eta^2}&=&-16\pi\left(T^{(m)}_{00}+{1\over 2}T^{(m)}\right)\;,\\
\label{scal}
\Box \phi-\frac{4}{\eta}{\partial \phi\over\partial\eta}&=&{6\chi\over\eta^2}\;,\\\label{asw}
\Box {\varphi}_{00}-\frac{4}{\eta}{\partial {\varphi}_{00}\over\partial\eta}&=&-16\pi T^{(m)}_{00}+{8\chi\over\eta^2}\;,\\\label{26q}
\Box {\varphi}_{0i}-\frac{4}{\eta}{\partial {\varphi}_{0i}\over\partial\eta}+\frac{4 {\varphi}_{0i}}{\eta^2}&=&-16\pi T^{(m)}_{0i}\;,
\\\label{27q}
\Box {\varphi}_{ij}-\frac{4}{\eta}{\partial {\varphi}_{ij}\over\partial\eta}&=&-16\pi T^{(m)}_{ij}\;,
\end{eqnarray}
where $\chi\equiv {\varphi}_{00}-\phi+\varphi/2$.

The residual gauge freedom is restricted by the equations
\begin{equation}
\label{rgf}
\Box \xi^0-\frac{4}{\eta}{\partial \xi^0\over\partial\eta}+\frac{4 \xi^0}{\eta^2}=0\;,
\qquad\quad\Box \xi^i-\frac{4}{\eta}{\partial \xi^i\over\partial\eta}=0\;,
\end{equation}
which can be derived from equation (\ref{gg}) after making use of equations (\ref{bbz}), (\ref{12}), and (\ref{12b}).
It is evident that the new gauge (\ref{gc}) allows us to make time transformations along with transformations of spatial coordinates using solutions of equations (\ref{rgf}). Thus, in the case of free gravitational waves one can construct an analog of the transverse-traceless (TT) gauge widely used for the description of physical effects of gravitational radiation in asympotically-flat space-times (see, e.g., \cite{ll}, \cite{wei}, \cite{6} as well as \cite{grish} and references therein).  
\section{Solutions of the linearized Einstein equations}\noindent
All differential equations given in the previous section have the following symbolic form
\begin{equation}
\label{s1}
\left(\Box-{a\over\eta}{\partial\over\partial\eta}-{b\over\eta^2}\right)F(\eta,{\bm{\it x}})=-4\pi S(\eta,{\bm{\it x}})\;,
\end{equation}
where $a$ and $b$ are constant numerical coefficients. This equation can be solved for arbitrary $a$ and $b$ using standard techniques \citep{7} and in the general case the solution is given in terms of the Fourier integrals based on Bessel functions with index $\nu=\sqrt{-b+(1-a)^2/4}$ that also defines the index of equation (\ref{s1}). 
In the specific case of the dust-dominated FLRW cosmological model the equations have three different indices $\nu=3/2$, $\nu=5/2$, and $\nu=7/2$. Fourier integrals constructed from the Bessel functions with such indices allow one to find solutions of equations (\ref{s1}) explicitly in terms of the retarded integrals. Without going into specific details of our calculations we give the final results:
\begin{enumerate}
\item General solution of equations (\ref{s1}) with index $\nu=3/2$:
\begin{eqnarray}
\label{pp2}
F(\eta,{\bm{\it x}})&=&{1\over\eta}{\partial\over\partial\eta}\left({\Psi\over\eta}\right)+{\cal F}_{3/2}\;,
\\
\label{pi}  
\Psi(\eta,{\bm{\it x}})&=&\int {d^3{\bm{\it x}}'\over |{\bm{\it x}}-{\bm{\it x}}'|}\left(\eta-|{\bm{\it x}}-{\bm{\it x}}'|\right)\int^{\eta-|{\bm{\it x}}-{\bm{\it x}}'|}_{\eta_0} uS(u,{\bm{\it x}}')du\;.
\end{eqnarray}
\item General solution of equations (\ref{s1}) with index $\nu=5/2$: 
\begin{eqnarray}
\label{pp1}
F(\eta,{\bm{\it x}})&=&{\partial\over\partial\eta}\left[{1\over\eta}{\partial\over\partial\eta}\left({\Psi\over\eta}\right)\right]+{\cal F}_{5/2}\;,
\\
\label{pi1}  
\Psi(\eta,{\bm{\it x}})&=&\int {d^3{\bm{\it x}}'\over |{\bm{\it x}}-{\bm{\it x}}'|}\left(\eta-|{\bm{\it x}}-{\bm{\it x}}'|\right)\int^{\eta-|{\bm{\it x}}-{\bm{\it x}}'|}_{\eta_0} udu\int^u_{u_0}S(v,{\bm{\it x}}')dv\;.
\end{eqnarray}
\item General solution of equations (\ref{s1}) with index $\nu=7/2$: 
\begin{eqnarray}
\label{pp3}
F(\eta,{\bm{\it x}})&=&\eta{\partial\over\partial\eta}\left\{ {1\over\eta}{\partial\over\partial\eta}\left[{1\over\eta}{\partial\over\partial\eta}\left({\Psi\over\eta}\right)\right]\right\}+{\cal F}_{7/2}\;,
\\
\label{pi3}  
\Psi(\eta,{\bm{\it x}})&=&\int {d^3{\bm{\it x}}'\over |{\bm{\it x}}-{\bm{\it x}}'|}\left(\eta-|{\bm{\it x}}-{\bm{\it x}}'|\right)\int^{\eta-|{\bm{\it x}}-{\bm{\it x}}'|}_{\eta_0} udu\int^u_{u_0} vdv\int^v_{v_0} S(w,{\bm{\it x}}'){dw\over w}\;.
\end{eqnarray}
\end{enumerate}
Here ${\cal F}_\nu(\eta,{\bm{\it x}})$ is a general solution of the homogeneous form of equation (\ref{s1}),
\begin{eqnarray}
\label{gf}
{\cal F}_\nu(\eta,{\bm{\it x}})={\rm Re}\int \hat{f}_\nu({\bm k})(k\eta)^{(1-a)/2}J_\nu(k\eta)\exp(i{\bm k}\cdot{\bm{\it x}})d^3{\bm k} \;,
\end{eqnarray}
where $J_\nu$ is the Bessel function of order $\nu$, $k=|{\bm k}|$, and $\hat{f}_\nu({\bm k})$ is the complex Fourier amplitude of ${\cal F}_\nu(\eta,{\bm{\it x}})$. It is important to realize that the homogeneous solution ${\cal F}_\nu(\eta,{\bm{\it x}})$ could in general involve free gravitational waves which would not be completely eliminated by  gauge transformations. 

\acknowledgments
This work has been supported by the Research Board Grant C-3-42847 of the University of Missouri-Columbia. The work of M. V. Sazhin has been partially supported by the RFFI grant 00-02-16350 (Russia). We thank A. L. Mellot, S. F. Shandarin, H. A. Feldman, N. Gnedin, A. A. Starobinsky, R. Juszkiewicz, and A. N. Petrov for fruitful discussions. M. V. Sazhin and J. Ramirez thank the Department of Physics and Astronomy, University of Missouri-Columbia, and S. M. Kopeikin thanks the Departamento de Fisica Teorica I, Universidad Complutense de Madrid, for hospitality. 

\end{document}